\documentclass[twoside,11pt]{article}

\usepackage{jmlr2e}
\usepackage[latin9]{inputenc}
\usepackage{array}
\usepackage{graphicx}
\usepackage{epstopdf}
\usepackage{array}
\usepackage{ragged2e}
\usepackage{tabularx}
\usepackage{booktabs}
\usepackage{amsmath}
\usepackage{algpseudocode}
\usepackage{tikz}
\usepackage{natbib}

\makeatletter

\let\textquotedbl="

\newcommand*\circled[1]{\tikz[baseline=(char.base)]{%
            \node[shape=circle,draw,inner sep=1pt] (char) {#1};}}

\jmlrheading{1}{2015}{1-48}{05/15}{}{Valerii Garnaga}

\makeatother

\ShortHeadings{Using of Neuro-Indexes}{Garnaga}
\firstpageno{1}

\begin{document}

\title{Using of Neuro-Indexes by Search Engines}

\author{\name Valerii Garnaga \email Valeriy.Garnaga@gmail.com \\
       \addr Information Technologies\\
       Kuban State University\\
       Krasnodar, 350040, Stavropolaskaya st, 149, Russia
}

\editor{}

\maketitle

\begin{abstract}%
The article describes a new data structure called neuro-index. It
is an alternative to well-known file indexes. The neuro-index is fundamentally
different because it stores weight coefficients in neural network.
It is not a reference type like \textquotedbl{}keyword\textendash{}position in a file\textquotedbl{}.
\end{abstract}


\section{Introduction}

Information is one of the most powerful resources available to the
organization. It is known that the ability to find the information
quickly and efficiently increases the size of profits and is an important
competitive advantage. The enterprise information infrastructure quality
can be judged on the effectiveness of the use of information. The
high-tech software allows you to quickly find the data required for
the business development, which is an attribute of a successful enterprise.

It is very interesting to improve an efficiency of obtaining data
relevant criteria of a user request using intelligent search and associative
semantic information based on artificially neural networks (ANN).
The effective neuro-index structure and the algorithm for the intelligent
information retrieval have established the researches in the areas
of artificial intelligent technologies and distributed computing environments.
The implemented prototype is based on the results of the researches.
The implementation is targeted for use in corporate networks, cloud
and distributed processing. It can reduce the cost of information
technology through the using of a new approach to intelligent information
retrieval.

It should be noted that the results of described technologies are
qualitatively different from the results obtained from the use of
fuzzy search algorithms. For instance, it is different from the based
on the Levenshtein distance \citep{levenshtein66}.

There are two types of file indexing:

1. Lexical indexing is designed to optimize logical requests.

2. Vector indexing allows queries by similarity.

At the late 80's Salton proposed a vector model \citep{salton89} as an alternative
to the lexical context-free indexing. Latent Semantic Indexing (LSI)
was developed in 1990 for verbal noise suppression and better response
relevancy \citep{deerwester90}. The model used singular value decomposition (SVD)
for the transition from a sparse matrix of words to a compact matrix
of principal eigenvalues \citep{golub89}. LSI showed considerable superiority
in the results \citep{dumais91} compared to the lexical method. However, the
complexity of the model often leads to significant loss in speed on
large collections compared with traditional logical techniques. Michael
Berry and Todd Letsche founded most workable systems based
on the LSI at Berkeley in 1995 \citep{letsche97}.

Essentially, these are two completely different paradigm indexing
\textendash{} Lexical indexing and Vector indexing. The only one thing
that unites them is the vector of occurrences of keywords in documents,
which is called ``classical index\textquotedblright .

\section{Materials and methods}

Let's show a difference between neuro-index and classical index \citep{dumais91}.
The classical index can be established by the vector shown below.

\begin{equation}\notag
\left(\begin{array}{l}
\left(k_{1},\ \left(p_{1\ 1},p_{1\ 2},\cdots,p_{1\ N_{1}}\right)\right)_{1},\\
\left(k_{2},\ \left(p_{2\ 1},p_{2\ 2},\cdots,p_{2\ N_{2}}\right)\right)_{2},\\
\cdots,\\
\left(k_{M},\ \left(p_{M\ 1},p_{M\ 2},\cdots,p_{M\ N_{M}}\right)\right)_{M}
\end{array}\right)
\end{equation}

Here $k_{i}$ are keywords and $p_{i\ j}$ are positions of keyword
$k_{i}$. The $M$ value is a number of keywords. The $N_{i}$ value
is a number of positions of keyword in a file. 

The neuro-index can be established by the matrix shown below.

\begin{equation}\label{eq:neuroindex}
W=\left[\begin{array}{ccc}
w_{1\ 1} & \cdots & w_{1\ I}\\
\vdots & \cdots & \vdots\\
w_{I\ 1} & \cdots & w_{I\ I}
\end{array}\right],\ i=1\cdots,I
\end{equation}

Here $W$ is a square matrix and $I$ is a number of neural network
neurons. The $w_{i\ j}$ is a weight coefficient from neuron number
$i$ to neuron number $j$. Also we can use zero values for $i$ and
$j$. This case will be described further in this article.

Let's see a mathematical model of used neural network. A
neuron can be established by the equation shown below \citep{galushkin02}.

\begin{equation}\label{eq:neuronmodel}
y=f\left(\sum_{n=1}^{N}v_{n}x_{n}+b\right)
\end{equation}

Here $(x_{1},x_{2},\ \cdots,x_{N})$ is an input vector for the
neuron. The vector $(v_{1},v_{2},\ \cdots,v_{N})$ consists of neuron
weight coefficients. The value of $N$ is a dimension of input signal.
The value of $b$ is an offset value. The $f()$ is an activation function for a non-linear
parameter transformation. The $y$ is an output value.

The neurons described in Equation \ref{eq:neuronmodel} can be united in the layers 
based on Equation \ref{eq:neuroindex}. The equation for neurons in layer is established by the next equation.

\begin{equation}\label{eq:layermodel}
y_{m_{j}}^{j}=f_{m_{j}}^{j}\left(\sum_{n_{j}=1}^{N_{j}}w_{m_{j}n_{j}}^{j}x_{n_{j}}^{j}+b_{m_{j}}^{j}\right),\ \ m_{j}=1,\ \cdots,M_{j},\ j=1,\ \cdots,J
\end{equation}

Here the value of $m$ is a number of neurons in a layer. The value
of $y_{m}$ is an output signal of the neuron number $m$ in a layer.
The value of $M$ is a number of neurons in a layer. The vector $x=(x_{1},x_{2},\ \cdots,x_{N})$
is an input signal for a layer. The matrix W$=\left\Vert {}w_{mn}\right\Vert {}$
is the neurons weight coefficients. The vector b$=(b_{1},b_{2},\ \cdots,b_{M})$
consists of offset values. The $f()$is an activation function for
the neuron number $m$in a layer.

The layers based on Equation \ref{eq:layermodel} are connected to a neural network. The number
of layers in the neural network is equal to $J$. The equation for
neuron number $m$ in the layer number $j$ is established by the next equation.

\begin{equation}\label{eq:networkmodel}
y_{m_{j}}^{j}=f_{m_{j}}^{j}\left(\sum_{n_{j}=1}^{N_{j}}w_{m_{j}n_{j}}^{j}x_{n_{j}}^{j}+b_{m_{j}}^{j}\right),\ \ m_{j}=1,\ \cdots,M_{j},\ j=1,\ \cdots,J
\end{equation}

The layers can be interconnected by connections of different types.
Let's assume that an input layer has number 0. In general, a signal
from an output layer number j will be sent to an input layer number
j+s. In this case we can establish three types of neural networks: 
\begin{enumerate}
\item s=1 \textendash{} feedforward neural network; 
\item s\textgreater1 \textendash{} neural network with cross-layer connections; 
\item s\textless1 \textendash{} recurrent neural network. 
\end{enumerate}
Input values for an input layer is a keyword, a keyword match number
and a degree of intelligence. They should be sent to an input of layer.
In this case using Equation \ref{eq:networkmodel} the $y_{m_{J}}^{J}$ will be output values
for a position in a file and a number of keyword matches.

I suggest using artificial neural networks for the search indexes
creation. For instance, the type of indexing artificial neural networks
(IANN) can be a feedforward neural network. Hidden layers are used
because they are necessary for a large volume of information processing.
Quantitative factors for neural network can be done in different ways.
It may be heuristic, genetic algorithms or an experimental selection
of parameters. In addition to keywords as input values of the neural
network there is a possibility of using the additional parameters.

A teaching process of IANN is based on the temporary stored classical
index. But I am going to change it in future. Now the classical index
should be generated using in-memory NoSQL DB. Back-propagation teaching
algorithm for IANN is used on the basis of the keywords and generated
classical index.

The input of IANN gets one keyword from the dictionary and the serial
number of occurrences of a keyword in a file. The output of the neural
network is a keyword occurrence position formed in a file. If the
occurrence of the specified sequence number does not exist then IANN
returns a non-existent value (for example -1).

I propose to take IANN initialized weights as an initial IANN for
training on any file. It allows recognizing the keywords regardless
of their form. Specially prepared initial IANN is used at the beginning
of the teaching process. It is trained on the set consisting of all
known keywords and on the keywords that vary according to the grammatical
rules. At the end of training (for indexing process) the IANN is checked
for grammar validity. This is another step in testing the resulting
values for the IANN.

Using the so-called ``any-time algorithm\textquotedblright{} \citep{boddy89}
will improve the search system efficiency. The essence of ``any-time
algorithm\textquotedblright{} is to gradually improve the partial
results. This capability is achieved through the use of neural network
technology.

The resulting weights of the IANN are index information for each file.
ANN Kohonen vector quantization is trained with a teacher \citep{kohonen89}.
It used to speed up the search process in the index information
according to the following steps:

\algrenewcommand{\alglinenumber}[1]{\scriptsize\circled{#1}}
\begin{algorithmic}[1]
\item Selection of the next observation is carried out after a file indexing.
A vector representing the inputs and outputs of the IANN makes the
observation.

\item Finding the best node on the ANN Kohonen map. It should be the
vector which weight is less than the weight of all different vectors.
This weight should be based on a minimum distance metric between the
terms and frequency of their occurrences and synonyms.

\item Determination of the number of neighbors and training and changing
weights of the vectors and its neighbors to bring them closer to the
observation.

\item Determination of the error map.

\item Re-training according to algorithm DLVQ Fundamentals \citep{stefano06}.
\end{algorithmic}

Let's describe the sample process of a search system based on neuro-indexes:

\algrenewcommand{\alglinenumber}[1]{\scriptsize\circled{#1}}
\begin{algorithmic}[1]
\item Search phrase is divided into pairs of terms. For example, the
phrase \textquotedbl{}what to buy for a holiday\textquotedbl{} can
be split into a pair of (what buy), (buy holiday).

\item Search priority to define pairs of corresponding direct order of
the search query (what buy), (buy holiday).

\item Using Kohonen ANN to look for a set of indices, which includes
the pairs.

\item Ranking the files according to the results of the IANS about the
proximity of keywords derived from the weights of the neuro-index
information with couples.
\end{algorithmic}

Search system teaching that is based on neuro-indexes requires more
computing resources than classical indexing process. This problem
can be solved by using of distributed computing. However, the search
process that is based on neuro-indexes is much faster than the search
process that is based on classical indexes and give us additional
advantages.

\section{Results}

Using the neuro-indexes instead of classical indexes demonstrates
a lot of innovative advantages. Bogdan Trofimov and I have implemented
the test-stand for comparison of these ideas. We have implemented
the text-search systems based on classical index and neuro-index and
compared them according to three criteria:

(i) storage size

\begin{center}
\includegraphics[width=1\textwidth,natwidth=412,natheight=62]{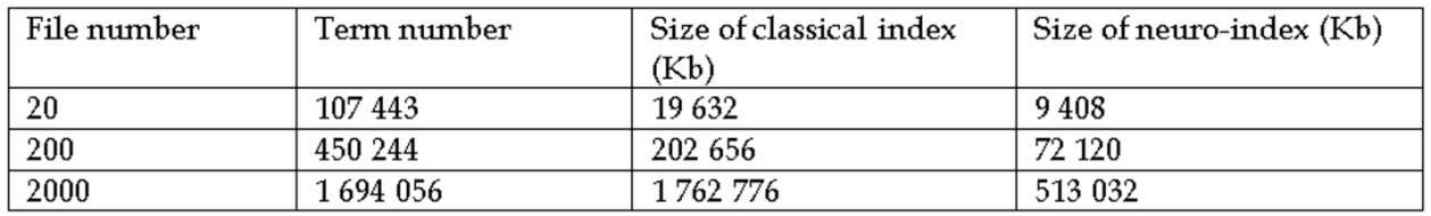} 
Table 1. Storage sizes for classical index and neuro-index.
\end{center}

The Table 1 shows that the sizes of indexes depend on term and file
numbers.

\begin{center}
\includegraphics[width=1\textwidth,natwidth=365,natheight=233]{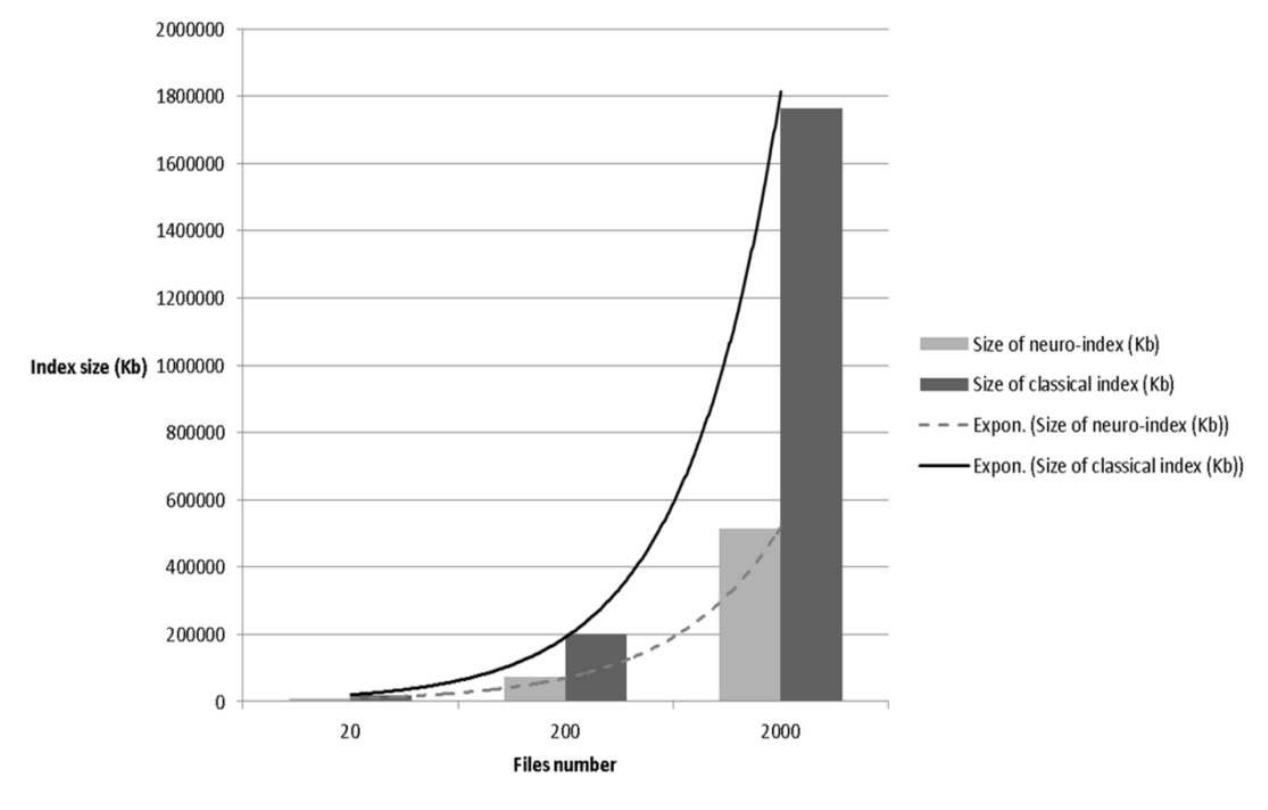} 
Figure 1. Storage sizes for classical index and neuro-index. 
\end{center}

Figure 1 shows that the size of classical index grows up dramatically
in comparison with the size of neuro-index.

(ii) search time

\begin{center}
\includegraphics[width=1\textwidth,natwidth=362,natheight=63]{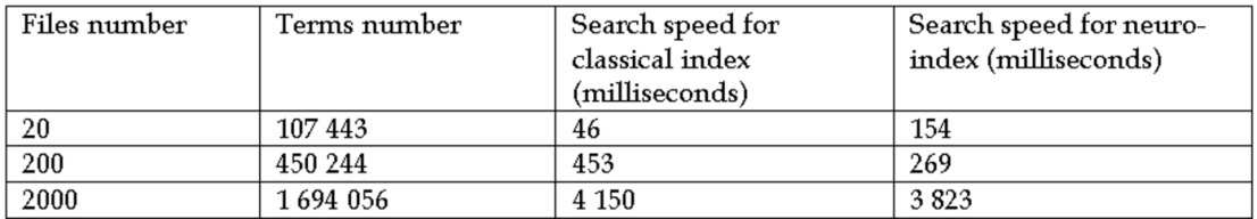} 
Table 2. Search speed for classical index and neuro-index. 
\end{center}

The Table 2 shows that the difference between classical index search
speed and neuro-index search speed is unambiguous but in big volumes
neuro-index search speed is faster than classical index search speed.

\begin{center}
\includegraphics[width=1\textwidth,natwidth=440,natheight=293]{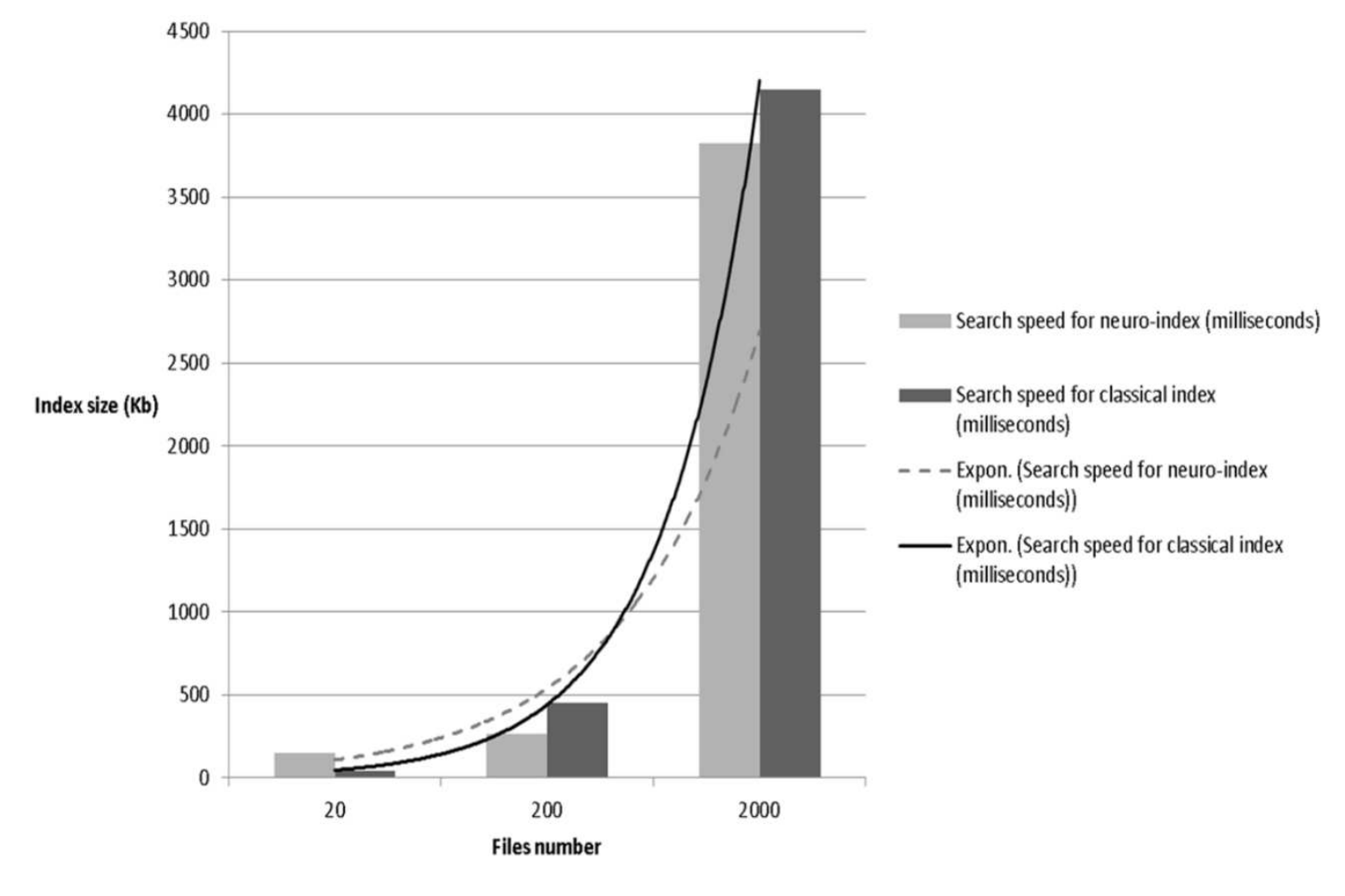}
Figure 2. Search speed for classical index and neuro-index.
\end{center}

The Figure 2 shows that search speed can be different. It depends on
search terms, source files and algorithm implementation. More effective
neuro-index algorithms and more accurate results about search speed
will be presented in the next article.

(iii) associative search

It is difficult for ordinary user to formulate query keywords especially
if there is a non-standard or restricted terminology. If he tries
to go beyond the narrow subject category he faces the problems of
synonymy and polysemy. Therefore it is very important to take into
account the context. The issues of the context representation can
be investigated with various approaches. This problem is effectively
solved by neural networks \citep{carpenter87}.

Implemented neuro-index search algorithm shows relevance factor for
exact match and for values that can have associative relations based
on ANN Kohonen map.

\section{Discussion}

Let's describe the possible ways of the proposed approach development. 
\begin{itemize}
\item Increase the speed of IANN teaching. 
\end{itemize}
Using of partially pre-trained neural networks \citep{Hansen90} on common
examples. This allows to train existing networks. The sample is given
in Figure 3.

\begin{center}
\includegraphics[width=1\textwidth,natwidth=385,natheight=255]{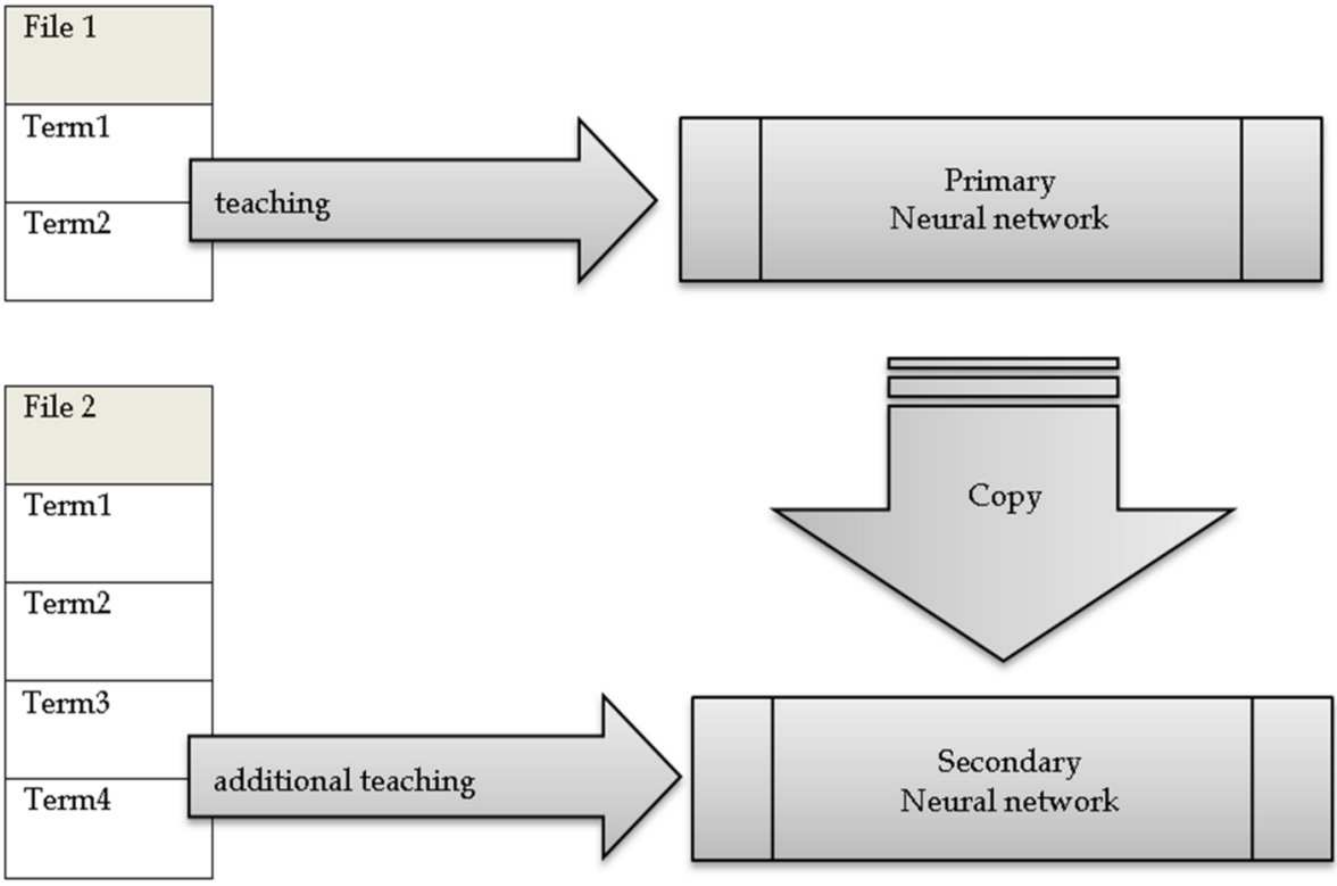}
Figure 3. Using of Primary Neural network.
\end{center}

Selection of the parameters and the structure of the neural network
according to the file type algorithms implementation. 
\begin{itemize}
\item Neural network topology selection. 
\end{itemize}
A network type is based on a problem type and data available for training.
Supervised teaching requires \textquotedbl{}expert\textquotedbl{}
assessment for each element of the sample. Sometimes getting such
assessment for a large data set is impossible. In this case the real
choice is teaching without a teacher. For example, it can be Kohonen
self-organizing map \citep{kohonen89} or Hopfield neural network \citep{hopfield82}.
For other problems solving (such as forecasting of time-series) expert
assessment is already contained in the input data and can be used
in their processing. 
\begin{itemize}
\item Selection of network characteristics. 
\end{itemize}
For perceptron neural network it can be a number of layers, a presence
or absence of bypass connections, a neurons activation functions.
The ability of the network to learn the higher is the larger connections
total number between neurons. But the number of connections is limited
to the top number of records in the training data. 
\begin{itemize}
\item Selection of teaching parameters. 
\end{itemize}
After selecting a specific topology it is necessary to selected neural
network training parameters. This step is especially important for
networks trained by a teacher. Network speed response convergence
to the correct answer depends on the correct choice of parameters.
Also the choice of a low teaching rate increases the convergence time
but sometimes it leads to network paralysis. Time training increasing
can either increase or decrease the convergence time because of the
surface shape error. According to the contradictory influence of parameters
we can conclude that their values should be chosen experimentally
based on the criterion of completion of training. It can be error
minimization or teaching time limit. 
\begin{itemize}
\item Preliminary assessment of information duplication from different files. 
\end{itemize}

\begin{center}
\includegraphics[width=1\textwidth,natwidth=373,natheight=265]{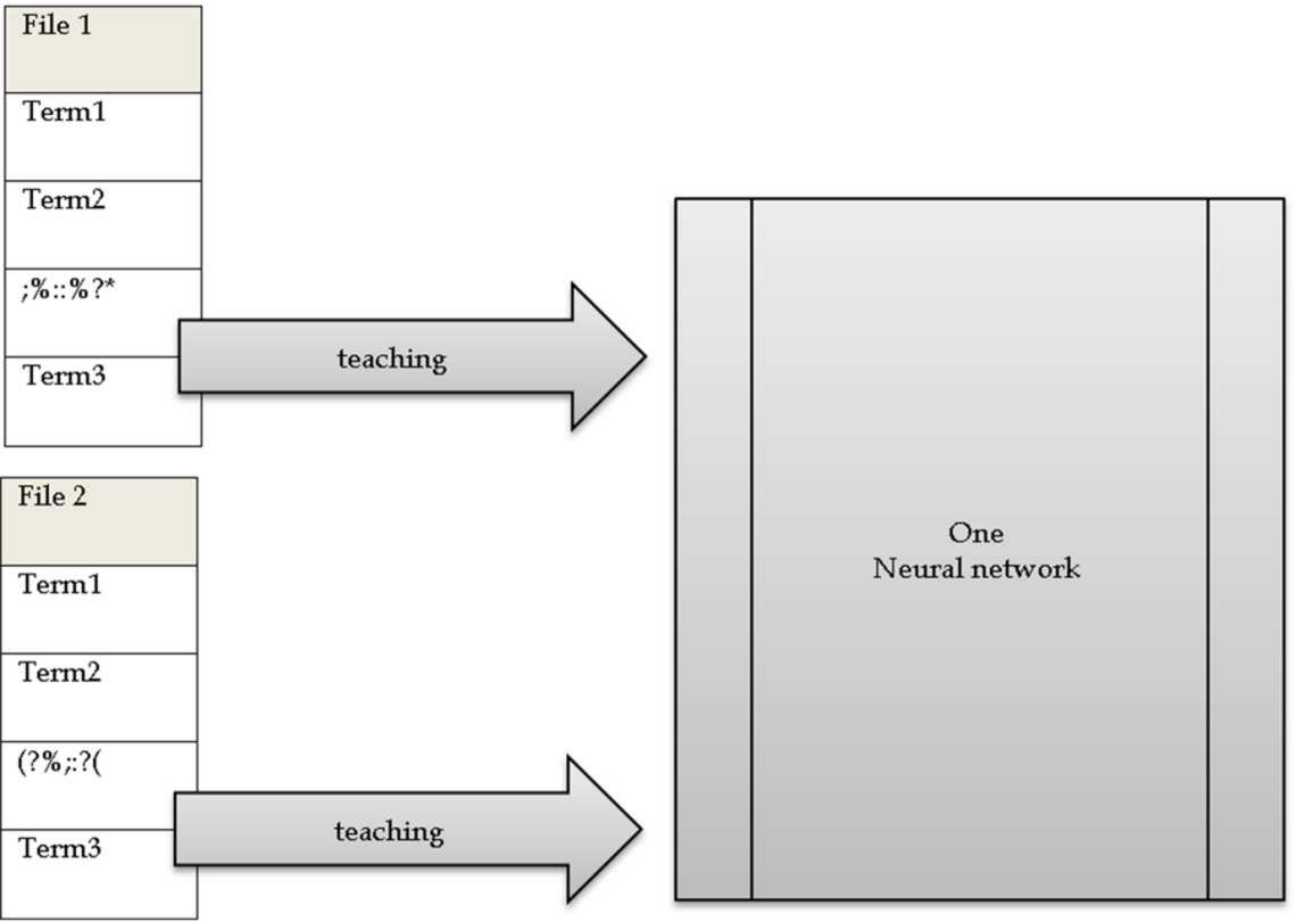}
Figure 4. Combination of Neural networks.
\end{center}

In this case it has been proposed to use the same neural network for
the files with the same text information as shown in Figure 4. 
\begin{itemize}
\item Exclusion of the classical indexing step from IANN teaching process. 
\end{itemize}
It has been suggested to implement an algorithm that allows training
the neural network using a stream of information from a file. This
approach requires a lot of additional research. However it directs
neuro-indexing methodology to a new level. 
\begin{itemize}
\item Research of special types of neural networks for using neuro-indexes. 
\end{itemize}
This research will definitely improve the effectiveness of the proposed
neural network search engine. There are many variants of neural network
types. These are the basic criteria for their classification: 
\begin{itemize}
\item types of neurons that make up the network; 
\item number of layers of neurons; 
\item direction transmission of signals; 
\item types of training samples; 
\item network purpose. 
\end{itemize}

\section*{Acknowledgments}
The author is very grateful to Yurii Koltsov, the dean of the faculty
of computer technologies and applied mathematics of Kuban State University,
and to Bogdan Trofimov, the postgraduate student, who works at Kuban
State University, for their assistance and feedback, which effected
the quality of the article. I also want to express my gratitude to
my colleague Valentina Semerzhidi for proofread this paper. This work
was supported in part by a grant from Russian Foundation for Basic
Research (RFBR) 13-01-00807 A.

\vskip 0.2in
\bibliography{garnaga2015}

\vskip 0.2in
Dr. Valerii V. Garnaga received PhD degree in 2004 and has worked as associated professor 
at the Kuban State University, Russia. His research interests are in artificial intelligence, 
distributed computing, intellectual search engines, cloud computing, distributed calculations, 
big data analysis. He has more than 15 publications in these areas and participated in number 
of research projects. Also, he was the chief of the project sponsored by Russian Foundation for Basic Research (RFBR).

\end{document}